# Correlation of ICME Magnetic Fields at Radially Aligned Spacecraft

**S.W. Good[1,2], R.J. Forsyth[1], J.P. Eastwood[1] and C. Möstl[3]**

[1]The Blackett Laboratory, Imperial College London, London, United Kingdom
[2]Department of Physics, University of Helsinki, Helsinki, Finland
[3]Space Research Institute, Austrian Academy of Sciences, Graz, Austria

Corresponding author e-mail: simon.good07@imperial.ac.uk

## Abstract

The magnetic field structures of two interplanetary coronal mass ejections (ICMEs) each observed by a pair of spacecraft close to radial alignment have been analysed. The ICMEs were observed *in situ* by MESSENGER and STEREO-B in November 2010 and November 2011, while the spacecraft were separated by more than 0.6 AU in heliocentric distance, less than 4° in heliographic longitude, and less than 7° in heliographic latitude. Both ICMEs took ~2 days to travel between the spacecraft. The ICME magnetic field profiles observed at MESSENGER have been mapped to the heliocentric distance of STEREO-B and compared directly to the profiles observed by STEREO-B. Figures that result from this mapping allow for easy qualitative assessment of similarity in the profiles. Macroscale features in the profiles that varied on timescales of 1 hour, and which corresponded to the underlying flux rope structure of the ICMEs, were well correlated in the solar east-west and north-south directed components, with Pearson's correlation coefficients of ~0.85 and ~0.95, respectively; microscale features with timescales of 1 minute were uncorrelated. Overall correlation values in the profiles of one ICME were increased when an apparent change in the flux rope axis direction between the observing spacecraft was taken into account. The high degree of similarity seen in the magnetic field profiles may be interpreted in two ways. If the spacecraft sampled the same region of each ICME (*i.e.* if the spacecraft angular separations are neglected), the similarity indicates that there was little evolution in the underlying structure of the sampled region during propagation. Alternatively, if the spacecraft observed different, nearby regions within the ICMEs, it indicates that there was spatial homogeneity across those different regions. The field structure similarity observed in these ICMEs points to the value of placing *in situ* space weather monitors well upstream of the Earth.

## 1. Introduction

Interplanetary coronal mass ejections (ICMEs) are discrete, large-scale magnetic field and plasma structures observed in the solar wind (*e.g.* Forsyth and Gosling, 2001). ICMEs variously display enhanced magnetic field strengths, depressed proton temperatures, bi-directional electron strahls, as well as a range of other *in situ* signatures (Zurbuchen and Richardson, 2006). Originating in the solar corona, ICMEs typically take 1-4 days to reach 1 AU, and form a direct link between the solar and terrestrial environments. Fast ICMEs with large and sustained southward magnetic field components are the primary drivers of adverse space weather at the Earth (Eastwood, 2008).

Sustained periods of geoeffective southward field are often associated with the subset of ICMEs that display a flux rope field geometry. Magnetic flux ropes consist of nested, helical field lines wound around a central axis, where the pitch angle of the field relative to the axis direction decreases as the axis is approached. A spacecraft passing through a flux rope will observe a smooth rotation of the magnetic field direction over a wide angle. ICMEs that display flux rope geometries and low plasma-$\beta$ are known as magnetic clouds (Burlaga *et al.*, 1981). It has been suggested that all ICMEs contain flux ropes, and that ICMEs observed without flux rope signatures are intersected by the observing spacecraft at the periphery of the ICME, away from the centrally-located flux rope (*e.g.* Russell *et al.*, 2005; Richardson and Cane, 2010). All current models of (I)CME initiation in the solar corona incorporate flux ropes, either as pre-existing structures or as by-products of the initiation process (Chen, 2011).

ICMEs may undergo non-radial deflections (Wang *et al.*, 2014), rotations (*e.g.* Nieves-Chinchilla *et al.*, 2013; Good *et al.*, 2015; Winslow *et al.*, 2016), and reconnective erosion (*e.g.* Ruffenach *et al.*, 2012) during propagation through interplanetary space. The prevalence of these effects within the inner heliosphere, and the degree to which they alter the underlying structure of ICMEs, remain open questions. In this study, the magnetic field structures of two ICMEs observed by a pair of spacecraft close to radial alignment with the Sun and separated by ~0.6 AU have been examined. These observations offer snapshots of the ICMEs at different stages in their propagation through the inner heliosphere. The extent to which the magnetic field structures of the ICMEs remained intact has been considered through direct comparison of the field profiles at each spacecraft. It has been found that the two ICMEs examined in this study displayed robust field structures, with at least one of the ICMEs showing evidence of rotation in its flux rope axis.

Both ICMEs displayed a flux rope structure, and both were magnetic clouds. The flux rope profiles have been mapped from the inner spacecraft to the heliocentric distances of the outer spacecraft through the application of a simple technique; the mapping has been performed in a way that factors out the radial expansion and drop in field magnitudes that occurred during propagation to allow direct comparison of the underlying field structure at the different observation points. Inputs required to perform the mapping include the spacecraft separation distance, the arrival times of the flux rope's leading and trailing edges at each spacecraft, and the magnetic field profiles observed at each spacecraft. Qualitative assessments of similarity in the magnetic field profiles are made from figures of overlapped data that result from the mapping, and similarity is quantified through the calculation of correlation coefficients for each field component.

In this work, we place particular emphasis on drawing conclusions directly from observations while making minimal assumptions about the ICMEs' global field structure. For example, no assumption is made as to whether the fields are force-free, and no particular cross-sectional shape for the flux ropes is assumed. Given the lack of plasma data at the inner spacecraft for both of the ICMEs studied, we do not attempt an analysis of the kind performed by Nakwacki *et al.* (2011), who considered, amongst other things, the evolution in magnetic flux, helicity, energy and expansion rate of a magnetic cloud observed *in situ* by radially aligned spacecraft at 1 and 5.4 AU.

We also note that this work does not provide any scheme for making predictions of arrival times, speeds or field magnitudes at an outer spacecraft based on *in situ* observations at an inner spacecraft; rather, these variables are obtained from the observations at both spacecraft *a posteriori* to produce best-fit mappings. This contrasts, for example, with the mapping technique recently introduced by Kubicka *et al.* (2016), who used inputs from one imager and an *in situ* spacecraft to predict ICME parameters at a second, radially aligned spacecraft.

In Section 2, the data mapping technique is described, and the result of its application to the two ICMEs are presented. An analysis of the overall correlation in the profiles, and an analysis of differences in correlation of features at microscopic and macroscopic scales, is also included in this section. A full discussion of the results and their interpretation is presented in Section 3.

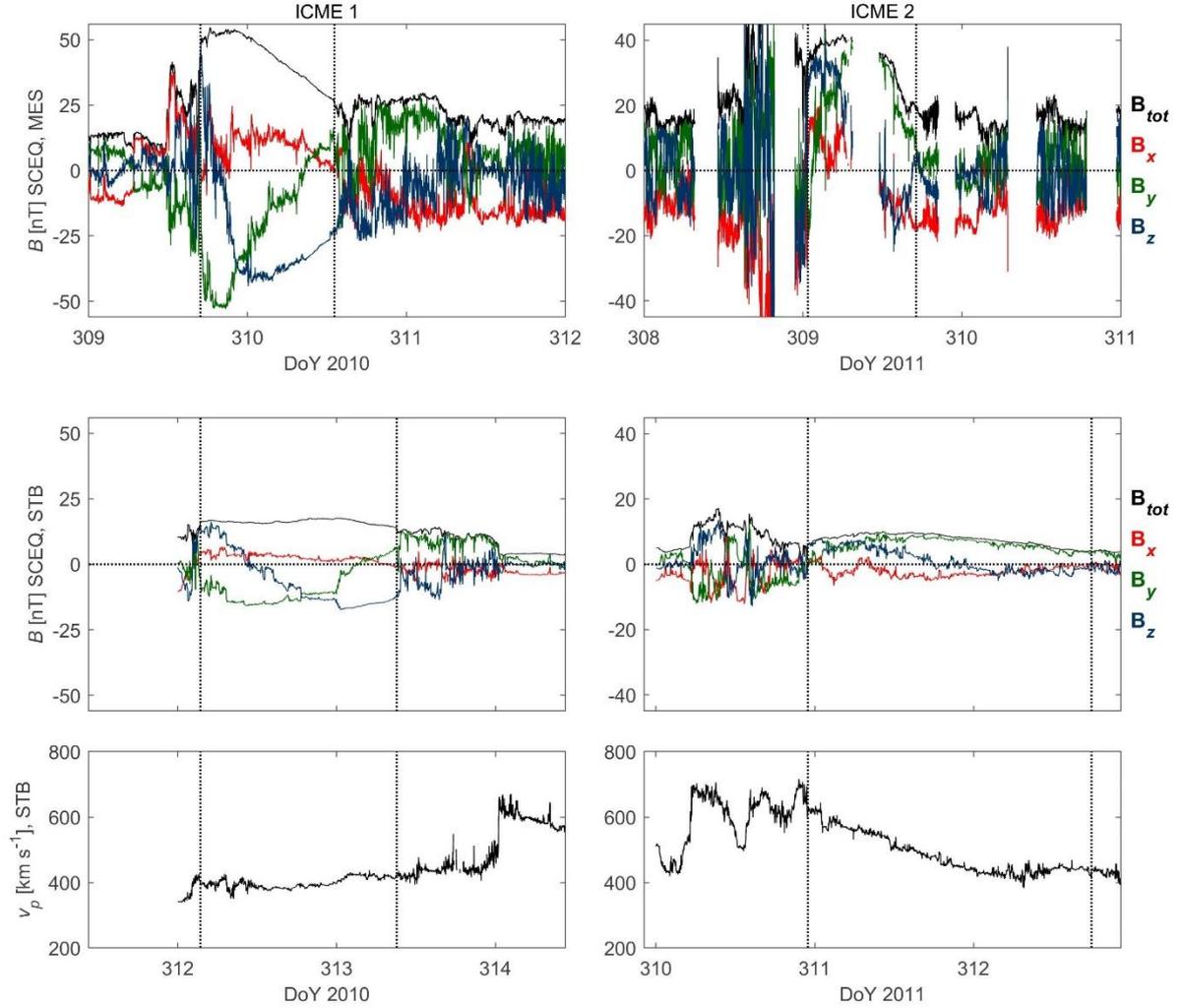

**Figure 1.** *In situ* data for ICME 1 (left panels) and ICME 2 (right panels). The panels show, from top to bottom, magnetic field data at MES, field data at STB, and the bulk plasma speed at STB. The black, red, green and blue lines in the field data panels correspond to the field magnitude, SCEQ $x$ component, SCEQ $y$ component and SCEQ $z$ component, respectively. All panels span a time period of 3 days for ease of comparison; the vertical axes in the field data panels have the same scaling for each ICME. Vertical dashed lines denote the flux rope boundaries. There was a data gap ahead of ICME 1's flux rope at STB.

## 2. Event Analysis

In this work, we examine the magnetic field structure of two ICMEs observed by a pair of radially aligned spacecraft. The two ICMEs have been selected for analysis because they both displayed flux rope structures that were clearly observed by each of the aligned spacecraft, because both ICMEs displayed relatively unambiguous boundaries, and because the observing spacecraft were reasonably well aligned. Also, the observing spacecraft were sufficiently well separated in radial distance for any evolution that may have occurred during propagation to become apparent. We note that the available plasma data indicates that both ICMEs were also magnetic clouds.

    The first ICME ("ICME 1") was observed by the *MErcury Surface, ENvironment GEochemistry, and Ranging* (MESSENGER; MES) spacecraft and the *Solar TErrestrial RElations*

*Observatory B* (STEREO-B; STB) in November 2010, while both spacecraft were separated by approximately 83° in heliographic longitude from the Sun-Earth line. This ICME was first described by Good and Forsyth (2016), and has recently been analysed by Amerstorfer *et al.* (2017) to test the ElEvoHI arrival time forecasting model. The flux rope of ICME 1 arrived at MES at 5 November 2010 16:52 UT; approximately 58 hours later, the flux rope arrived at STB. During the observation period, the spacecraft were separated by an average of 7.0° in heliographic latitude, 1.0° in heliographic longitude, and 0.618 AU in radial distance. The left-hand panels in Figure 1 show, from top to bottom, the magnetic field magnitude and components at MES, the field magnitude and components at STB, and the bulk plasma speed profile at STB. Vertical dashed lines denote the flux rope boundaries. The three panels all display data across a time span of 3 days. Magnetic field data are displayed in Spacecraft Equatorial (SCEQ) coordinates, in which $z$ is parallel to the solar rotation axis, $y$ points to solar west, and $x$ completes the right-handed system. SCEQ coordinates are very similar to the RTN system for spacecraft near the solar equatorial plane, and identical to it for a spacecraft in the plane.

The second ICME analysed ("ICME 2") was also observed by MES and STB during radial alignment, one year to the day after the passage of ICME 1. ICME 2's flux rope arrived at MES at 5 November 2011 00:43 UT and at STB at 6 November 2011 22:57 UT; the spacecraft were separated by an average of 6.8° in latitude, 3.5° in longitude and 0.647 AU in radial distance during the observation period. At that time, the aligned spacecraft were separated by ~103° in longitude from the Sun-Earth line. The right-hand panels in Figure 1 display the magnetic field data at MES and STB and the bulk plasma speed at STB for the ICME. Key parameters for both ICMEs analysed are listed in Table 1.

Data from magnetometers on board MES (MAG; Anderson *et al.*, 2007) and STB (IMPACT MAG; Acuña *et al.*, 2007) are used in this study. MES data were obtained from the PDS:PPI archive (`pds-ppi.igpp.ucla.edu`), and STB data from the SPDF CDAWeb archive (`cdaweb.sci.gsfc.nasa.gov`). Only magnetic field data were routinely available at MES since the spacecraft did not carry any dedicated instruments for analysing the solar wind plasma; both magnetic field and plasma data were available at STB. Data in RTN coordinates were obtained from the archives and transformed to the SCEQ system through a rotation about $T$ by the heliographic latitude of the spacecraft at the observation time, such that $N$ became aligned with the solar rotation axis direction.

|  | ICME 1 | ICME 2 |
|---|---|---|
| *Kinematics* | | |
| $\langle r_{\mathrm{H}} \rangle$, MES [AU] | 0.465 | 0.439 |
| $r_{\mathrm{H}}$, STB [AU] | 1.083 | 1.086 |
| $\Delta\theta_{\mathrm{HGI}}$, $\Delta\varphi_{\mathrm{HGI}}$ | 7.0°, 1.0° | 6.8°, 3.5° |
| $t_{\mathrm{L}}$, MES | 5 Nov 2010 16:52 UT | 5 Nov 2011 00:43 UT |
| $t_{\mathrm{T}}$, MES | 6 Nov 2010 13:08 UT | 5 Nov 2011 17:05 UT |
| $\Delta t_{\mathrm{MES}}$ | 20 hr 16 min | 16 hr 22 min |
| $t_{\mathrm{L}}$, STB | 8 Nov 2010 03:24 UT | 6 Nov 2011 22:57 UT |
| $t_{\mathrm{T}}$, STB | 9 Nov 2010 09:04 UT | 8 Nov 2011 17:48 UT |
| $\Delta t_{\mathrm{STB}}$ | 29 hr 40 min | 42 hr 51 min |
| $\Delta t_{\mathrm{L}}$ | 58 hr 32 min | 46 hr 14 min |
| $\Delta t_{\mathrm{T}}$ | 67 hr 56 min | 72 hr 43 min |
| $v_{\mathrm{L}}$, STB [km s$^{-1}$] | 402 | 618 |
| $v_{\mathrm{T}}$, STB [km s$^{-1}$] | 418 | 410 |
| $v_{\mathrm{EXP}}$, STB [km s$^{-1}$] | -16 | 208 |
| $\langle v_{\mathrm{L}} \rangle$ [km s$^{-1}$] | 437 | 580 |
| $\langle v_{\mathrm{T}} \rangle$ [km s$^{-1}$] | 380 | 370 |
| $\langle v_{\mathrm{EXP}} \rangle$ [km s$^{-1}$] | 57 | 210 |
| $v_{\mathrm{c}}$, STB [km s$^{-1}$] | 394 | 473 |
| $\langle v_{\mathrm{c}} \rangle$ [km s$^{-1}$] | 409 | 475 |
| $\langle B_{\mathrm{STB}}/B_{\mathrm{MES}} \rangle$ | 0.39 | 0.23 |
| *Flux rope axis* | | |
| Axis direction, MES | $\theta_{\mathrm{A}} = -41°$, $\varphi_{\mathrm{A}} = 287°$ | - |
| Axis direction, STB | $\theta_{\mathrm{A}} = -20°$, $\varphi_{\mathrm{A}} = 271°$ | - |
| Axis separation | $\psi = 19°$ | - |
| $\lambda_1/\lambda_2$, $\lambda_3/\lambda_2$, MES | 4.70, 0.14 | - |
| $\lambda_1/\lambda_2$, $\lambda_3/\lambda_2$, STB | 4.13, 0.03 | - |
| *Correlation* | | |
| $C$ | [-0.04, 0.82, 0.91] | [0.12, 0.70, 0.91] |
| $C$, axis-aligned | [-0.31, 0.89, 0.96] | - |
| $C_{\mathrm{L}}$ | [-0.02, 0.85, 0.92] | [0.14, 0.84, 0.94] |
| $C_{\mathrm{S}}$ | [-0.06, 0.05, 0.13] | [0.03, 0.02, -0.07] |
| $C_{\mathrm{L}}$, axis-aligned | [-0.41, 0.94, 0.96] | - |
| $C_{\mathrm{S}}$, axis-aligned | [-0.09, 0.06, 0.12] | - |

**Table 1.** Key parameters of the two ICMEs analysed. $\langle r_{\mathrm{H}} \rangle$ is the mean heliocentric spacecraft distance, $\Delta\theta_{\mathrm{HGI}}$ and $\Delta\varphi_{\mathrm{HGI}}$ are the latitudinal and longitudinal spacecraft separations, $t_{\mathrm{L}}$ and $t_{\mathrm{T}}$ are the observation times of the flux rope leading and trailing edges, $\Delta t_{\mathrm{MES}}$ and $\Delta t_{\mathrm{STB}}$ are the ICME crossing times at each spacecraft, $\Delta t_{\mathrm{L}}$ and $\Delta t_{\mathrm{T}}$ are the leading and trailing edge propagation times, $v_{\mathrm{L}}$ and $v_{\mathrm{T}}$ are the leading and trailing edge plasma speeds at STB, $v_{\mathrm{EXP}}$ is the expansion speed at STB, $\langle v_{\mathrm{L}} \rangle$ and $\langle v_{\mathrm{T}} \rangle$ are the mean speeds of the leading and trailing edges during propagation, $\langle v_{\mathrm{EXP}} \rangle$ is the mean expansion speed during propagation, $v_{\mathrm{c}}$ is the cruise speed at STB, $\langle v_{\mathrm{c}} \rangle$ is the mean cruise speed during propagation, $\langle B_{\mathrm{STB}}/B_{\mathrm{MES}} \rangle$ is the mean ratio of the field magnitude at the outer to the inner spacecraft, $\theta_{\mathrm{A}}$ and $\varphi_{\mathrm{A}}$ the latitude and longitude directions of the flux rope axes, $\lambda_1/\lambda_2$ and $\lambda_3/\lambda_2$ are the ratios of the maximum and minimum eigenvalues to the intermediate values found in the MVA, $C$ is the overall correlation coefficient, $C_{\mathrm{L}}$ the macroscale correlation, and $C_{\mathrm{S}}$ the microscale correlation.

## 2.1. Radial Alignment Mapping

In order to compare directly the magnetic field profiles of the ICMEs at each spacecraft, the magnetic field measurements within the flux rope at MES have been mapped forward in time and heliocentric distance to overlap with the measurements made at STB. Conceptually, we imagine each magnetic field vector measured at the inner spacecraft being frozen-in to a discrete plasma parcel. The collection of parcels that constitute the ICME propagate radially to the heliocentric distance of the outer spacecraft. The mapping involves determining the arrival time of each parcel and its magnetic field vector at the outer spacecraft distance, which requires knowledge of the parcel speeds. It is assumed that the parcel velocities are entirely in the radial direction.

ICMEs tend to expand in radial width as they propagate from the Sun until reaching some equilibrium state with the ambient solar wind. The speed observed *in situ* at the leading edge of an expanding ICME will exceed the speed at the trailing edge, and will typically decline linearly in between. For both ICMEs studied, measurements of the speed profiles were only available at the outer spacecraft. ICME 1 displayed a flat speed profile at STB (bottom-left panel, Figure 1), indicating that radial expansion had ceased by the time it arrived at the spacecraft. In contrast, ICME 2 displayed a linearly declining speed profile (bottom-right panel, Figure 1), indicating that the ICME was still expanding. These profiles represent a series of instantaneous speeds measured as the ICMEs passed over the outer spacecraft, and may be different to the speeds within the ICMEs during propagation.

To perform the mapping, we directly calculate the mean propagation speeds of the leading and trailing edges from their arrival times at each spacecraft, and assume a linear decline in speed between these values. The resulting speed profile represents the mean speed profile held by the ICME during propagation between the spacecraft. The mean speeds of the leading and trailing edges are simply given by

$$\langle v_{\mathrm{L}} \rangle = R_{\mathrm{L}}/(t_{\mathrm{L2}} - t_{\mathrm{L1}}) \equiv R_{\mathrm{L}}/\Delta t_{\mathrm{L}} \qquad (1a)$$

and

$$\langle v_{\mathrm{T}} \rangle = R_{\mathrm{T}}/(t_{\mathrm{T2}} - t_{\mathrm{T1}}) \equiv R_{\mathrm{T}}/\Delta t_{\mathrm{T}} \qquad (1b)$$

respectively, where $t_L$ is the leading edge arrival time, $t_T$ is the trailing edge arrival time, $R_{\mathrm{L}}$ and $R_{\mathrm{T}}$ are the propagation distances of the leading and trailing edges, respectively, subscript 1 denotes the inner spacecraft, 2 the outer spacecraft, $\Delta t_{\mathrm{L}} = (t_{\mathrm{L2}} - t_{\mathrm{L1}})$, and $\Delta t_T = (t_{\mathrm{T2}} - t_{\mathrm{T1}})$. It may be shown that the mean speed profile is given by

$$v = \frac{\langle v_{\mathrm{T}} \rangle - \langle v_{\mathrm{L}} \rangle}{t_{\mathrm{T1}} - t_{\mathrm{L1}}} (t_1 - t_{\mathrm{L1}}) + \langle v_{\mathrm{L}} \rangle . \qquad (2)$$

A parcel observed at time $t_1$ at the inner spacecraft will have a mean propagation speed between the inner and outer spacecraft given by $v$. The arrival time of this parcel at the outer spacecraft is thus given by

$$t_2 = t_1 + \frac{R}{v}, \qquad (3)$$

where $R$ is the propagation distance of the parcel between the spacecraft. Modelling the speed profile in this way constrains the leading and trailing edges at the inner spacecraft to map to the times at which they were observed at the outer spacecraft. Magnetic field vectors between the edges are not constrained to overlap with any particular feature observed at the outer spacecraft.

The gradient in the speed profile produces a non-self-similar mapping from one spacecraft to the other. Figure 2 demonstrates the effect of a sloping speed profile over time on an arbitrary, initially symmetric magnetic field time series. The left-hand side of the figure shows the initial field profile and the right-hand side shows the profile at some later time at another spacecraft. Faster-moving features that appear earlier in the series appear relatively closer together at the second spacecraft, while slower-moving features later in the series appear relatively spread out. This effect is present in various models of ICME evolution (*e.g.* Osherovich *et al.*, 1993). Note that Figure 2 illustrates the relative spacing of features on arrival at the second spacecraft, and not any changes in magnitude that may have occurred during propagation.

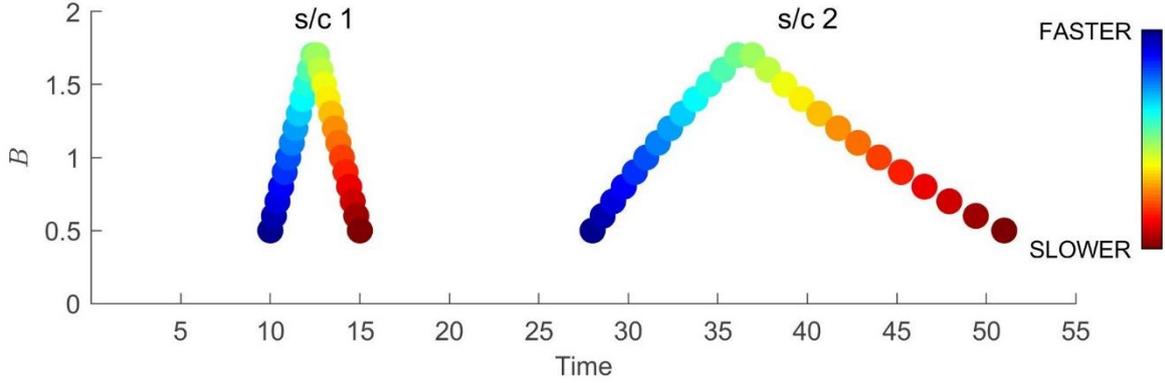

**Figure 2.** The distortion of an initially symmetric, arbitrary, normalised magnetic field feature propagating with a linearly declining speed profile. Each point corresponds to a separate plasma parcel travelling from s/c 1 to s/c 2, where the mean speed of each parcel between the spacecraft is denoted by its colouring. The leading edge of the feature travels at twice the mean speed of the trailing edge in this example.

## 2.2. Mapping of ICME 1 and ICME 2 Data

The left-hand side of Figure 3 displays the mapping of ICME 1's magnetic field measurements at MES to the location of STB. The pale-coloured lines show the mapped MES data overlaying the dark-coloured lines of the STB data. MES data at a time resolution of 1 minute was used to perform the mapping, and the STB data displayed in the figure is also at a 1 minute resolution. The panels on the left-hand side of Figure 3 show, from top to bottom, the three SCEQ components of the magnetic field, $\hat{B}_x$, $\hat{B}_y$ and $\hat{B}_z$, the latitude angle of the field direction, $\theta_B$, and the angle between the projection of the field vector onto the $x$-$y$ plane and the $x$ direction, $\varphi_B$. The field components have been normalised to the field magnitude. The vertical dashed lines mark the leading and trailing edges of the ICME's flux rope. Qualitatively, it can be seen in Figure 3 that there are significant similarities between the STB and mapped MES profiles for all three field components and the field direction angles. Note that the distortions illustrated by Figure 2 were very minor in the mapped MES data for ICME 1; greater distortion would have arisen if the speed profile gradient had been steeper, or if the spacecraft separation distance had been larger.

The flux rope orientation at MES was different to the orientation observed at STB for ICME 1. The orientation of a flux rope's central axis may be estimated through minimum variance analysis (MVA). MVA is a widely used technique that involves finding the eigenvalues and eigenvectors of the covariance matrix of the field data within the flux rope, where the eigenvector associated with the intermediate eigenvalue will correspond ideally to the direction of the rope axis (Goldstein, 1983). MVA gives an estimated orientation of [$\theta_A = -41°$, $\varphi_A = 287°$] for the mapped ICME 1 MES data and an orientation of [$\theta_A = -27°$, $\varphi_A = 271°$] at STB, where $\theta_A$ and $\varphi_A$ are analogous to the field direction angles defined above. Visual inspection of the data suggests an intermediate variance direction close to the $-y$ direction (solar east) at both spacecraft, in agreement with the MVA. The direction angles found in the mapped MES data are the same as those found in the original data series, to the nearest degree. For the mapped MES data, the ratio of the minimum to intermediate eigenvalue, $\lambda_1/\lambda_2$, was equal to 4.70, and the ratio of the maximum to intermediate value, $\lambda_3/\lambda_2$, was 0.14. The corresponding values at STB were 4.13 and 0.03, respectively. The ratios in both datasets meet the Siscoe and Suey (1972) conditions, namely $\lambda_1/\lambda_2 > 1.37$ and $\lambda_3/\lambda_2 < 0.72$, indicating that the variance directions were well defined.

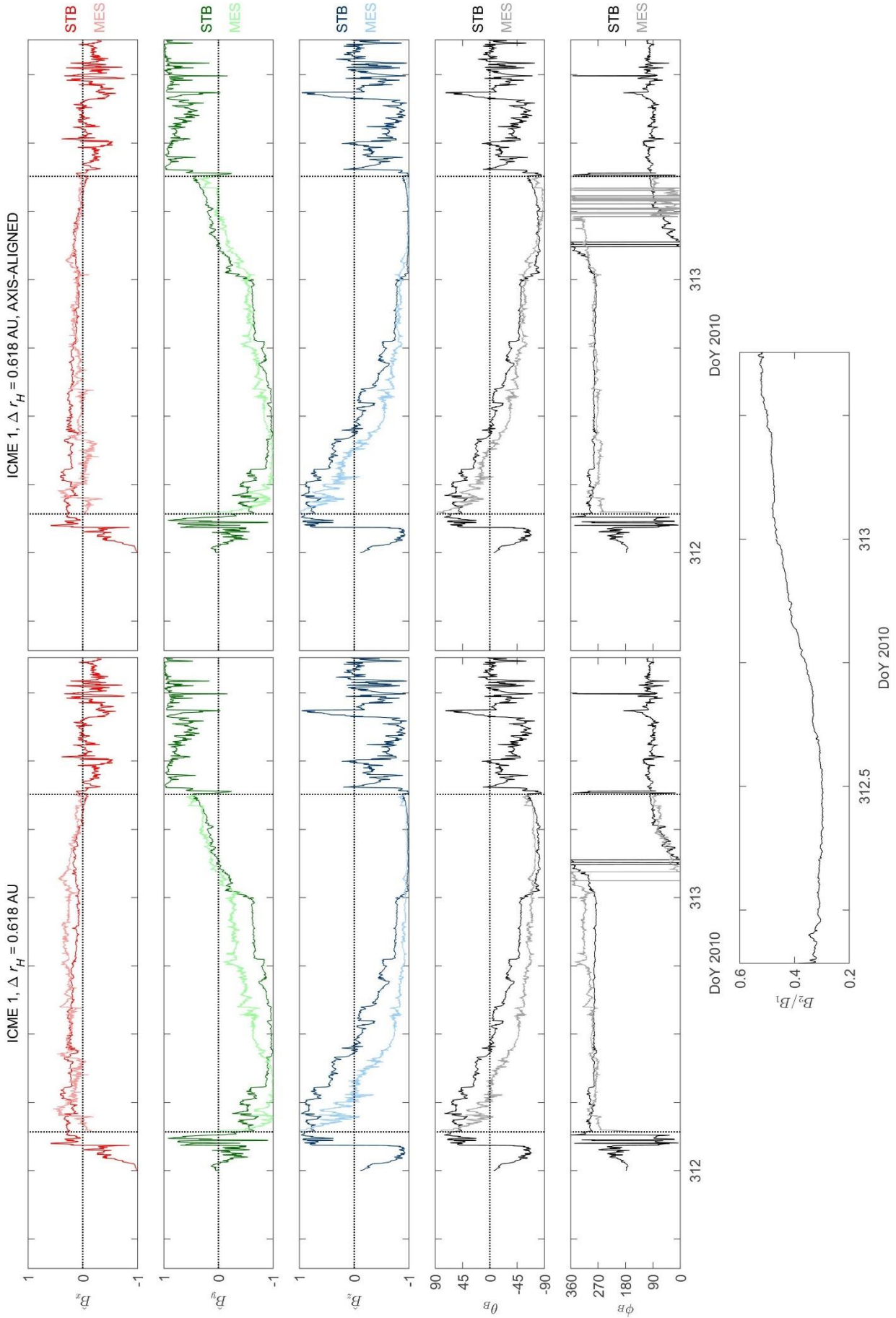

**Figure 3.** (*previous page*) ICME 1 mappings. Paler-coloured lines show the mapped MES flux rope data, overlaying the darker-coloured lines of the STB data. The panels show the three normalised magnetic field components in SCEQ co-ordinates, $\hat{B}_x$, $\hat{B}_y$ and $\hat{B}_z$, and the field direction angles, $\theta_B$ and $\varphi_B$ (see text for details). Vertical dashed lines denote the flux rope boundaries. The left-hand panels show the initial mapping using Equations 2 and 3, and the right-hand panels show the axis-aligned mapping. The bottom centred panel shows the ratio of the STB to MES field magnitude throughout the flux rope.

We now consider the effect of transforming the mapped MES flux rope to align its central axis, pointing in direction $\mathbf{a_{MES}}$, with the axis found in the STB rope data, pointing in direction $\mathbf{a_{STB}}$. This transformation has been achieved by defining the plane in which both axis directions are coplanar, determining the angle between the axes in that plane, $\psi = \tan^{-1}(|\mathbf{a_{MES}} \times \mathbf{a_{STB}}| / \mathbf{a_{MES}} \cdot \mathbf{a_{STB}})$, and determining the direction normal to that plane, $\mathbf{A} = \mathbf{a_{MES}} \times \mathbf{a_{STB}}$; rotating all of the magnetic field vectors of the mapped MES data by $\psi$ about the normal direction $\mathbf{A}$ such that the flux rope axes become aligned gives the required transformation. This transformation cancels out the apparent change in axis orientation between the spacecraft, and cancels out differences between the profiles that are purely due to the different orientations of the flux rope.

This transformation of the mapped MES data for ICME 1 is shown on the right-hand side of Figure 3, overlaying the STB rope data. Compared to the left-hand side, there is a marked increase in similarity of the $\hat{B}_y$ and $\hat{B}_z$ profiles, particularly in the front and middle regions of the rope. There is little overall change in similarity for $\hat{B}_x$. The axis separation, $\psi$, was equal to 19°.

The mapping described by Equation 2 and 3 has also been applied to ICME 2. The result of this mapping is displayed in Figure 4. The panels in the figure have a corresponding layout to those in Figure 3. There is good qualitative agreement in the $\hat{B}_z$ profiles and the field direction angles, marginally less agreement in the $\hat{B}_y$ profiles, and considerably less agreement in $\hat{B}_x$.

There is a significant gap in the MES data for ICME 2 during which the spacecraft was passing through Mercury's magnetosphere. The gap spans 30% of the flux rope time series, and obscures the spacecraft's closest approach to the flux rope axis. These factors are the likely cause of the dubious estimate of the axis orientation obtained from MVA when applied to the normalised field data. We therefore have not attempted the axis-aligned mapping for ICME 2 that was performed for ICME 1. Good *et al.* (2015) provide a more detailed analysis of this ICME's flux rope orientation.

The bottom-centre panels in Figures 2 and 3 show the ratio of the field magnitude at the outer spacecraft to the magnitude at the inner spacecraft for ICMEs 1 and 2, respectively, across the flux rope profiles. Values below unity indicate a drop in field magnitude. In ICME 1, the ratio rose smoothly from ~0.3 in the front half of the rope to ~0.5 towards the trailing edge. The mean value of the magnitude ratio across the rope was 0.39. In contrast, the ratio for ICME 2 was relatively flat across the profile, with a mean value of 0.23.

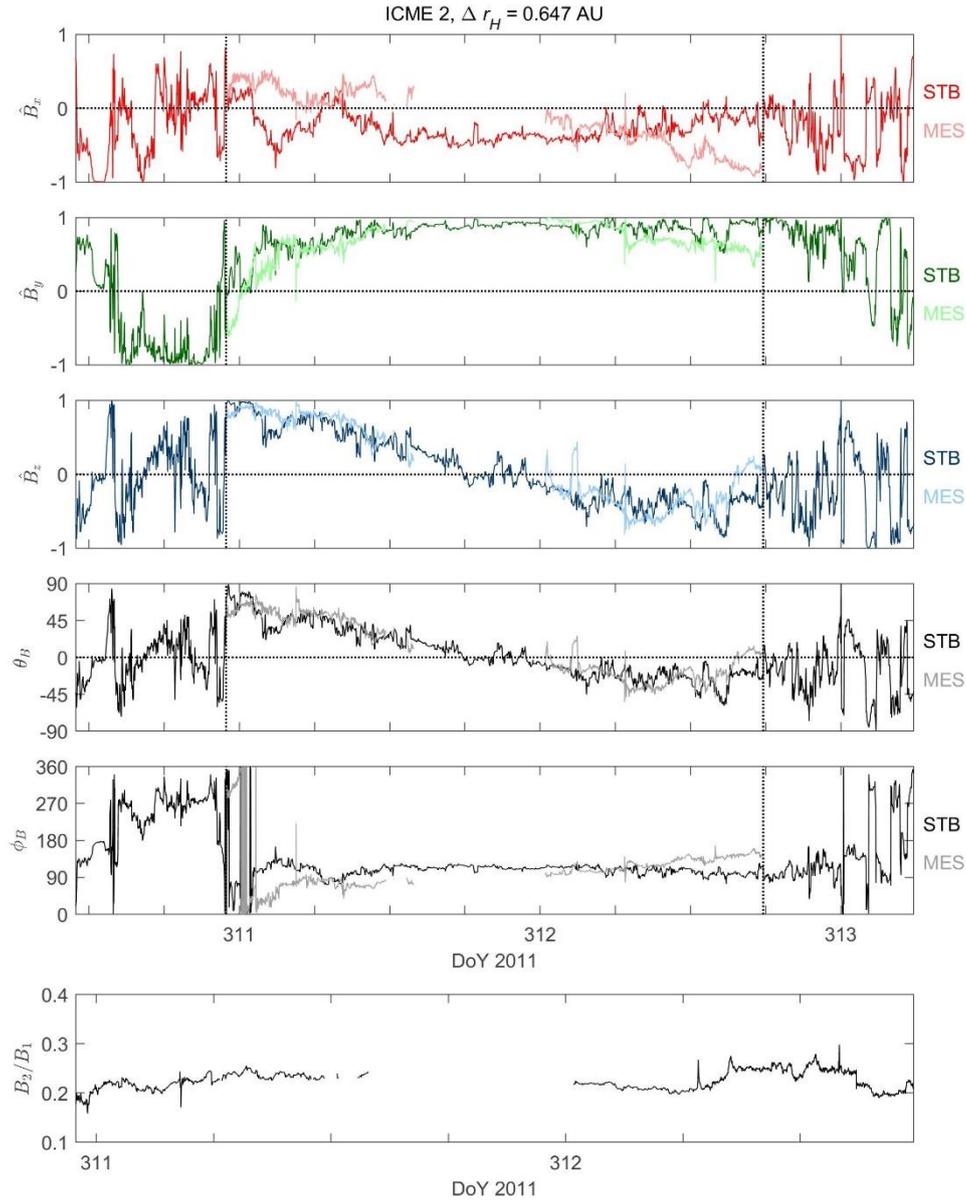

**Figure 4.** ICME 2 mapping. The figure is presented in the same way as Figure 3.

### 2.2.1. ICME Expansion Speeds and Crossing Times

The expansion speed of an ICME, $v_{EXP}$, may be defined as the difference between its leading and trailing edge speeds, $v_L - v_T$. The expansion speeds observed *in situ* at STB were -16 km s$^{-1}$ for ICME 1 and 208 km s$^{-1}$ for ICME 2. These values, and the leading and trailing-edge plasma speeds from which they are calculated, are listed in Table 1. The $v_{EXP}$ values indicate that ICME 1 had ceased to expand by the time it arrived at STB and that ICME 2, in contrast, was still rapidly expanding at STB. $v_{EXP}$ values could not be calculated at MES given the lack of plasma data.

The mean expansion speeds during propagation between the spacecraft, $\langle v_{EXP} \rangle = \langle v_L \rangle - \langle v_T \rangle$, were 57 km s$^{-1}$ for ICME 1 and 210 km s$^{-1}$ for ICME 2. The mean propagation speeds of the leading edge, $\langle v_L \rangle$, and trailing edge, $\langle v_T \rangle$, are as defined by Equation 1. The positive, non-zero $\langle v_{EXP} \rangle$ values confirm that both ICMEs expanded during propagation, and account for the increased ICME crossing times, $t_L - t_T$, seen at STB; for ICME 1, the crossing time rose from 20 hr 16 min at MES to 29 hr 40 min at STB, and from 16 hr 22 min at MES to 42 hr 51 min at STB for ICME 2. The mean expansion speed of ICME 2 between MES and STB was very similar to the expansion speed observed at STB, indicating that the ICME's expansion speed was roughly constant during propagation between the spacecraft.

The ICME crossing time at STB, $\Delta t_{STB}$, may be related to $\langle v_{EXP} \rangle$, $\langle v_L \rangle$, $\langle v_T \rangle$ and the crossing time at MES, $\Delta t_{MES}$. Noting that $\Delta t_{STB} - \Delta t_{MES} = \Delta t_T - \Delta t_L$, where $\Delta t_L$ and $\Delta t_T$ are the leading and trailing edge propagation times, respectively, and by approximating $R_L \approx R_T \equiv R$ such that $\langle v_L \rangle = R/\Delta t_L$ and $\langle v_T \rangle = R/\Delta t_T$, it may be shown that

$$\Delta t_{STB} \approx \Delta t_{MES} + R \frac{\langle v_{EXP} \rangle}{\langle v_L \rangle \langle v_T \rangle} . \qquad (4)$$

Equation 4 indicates simply that the ICME crossing time at the outer spacecraft (STB) is equal to the crossing time at the inner spacecraft (MES) plus a term due to the ICME's expansion (or contraction, if $\langle v_{EXP} \rangle$ were negative); the approximate equation becomes precisely equal when $R_L = R_T$. This relation is in agreement with the values quoted above for both ICMEs.

For ICME 1, the *in situ* speed of the leading edge at STB, $v_L$, was approximately 402 km s$^{-1}$, lower than the $\langle v_L \rangle$ value of 437 km s$^{-1}$. In contrast, the in situ trailing edge speed, $v_T$, was around 418 km s$^{-1}$, higher than the $\langle v_T \rangle$ value of 380 km s$^{-1}$. Thus, the leading edge decelerated, the trailing edge accelerated, and the speed profile across the rope flattened during propagation. For ICME 2, both the leading and trailing edge speeds at STB (618 km s$^{-1}$ and 410 km s$^{-1}$, respectively) were somewhat higher than their mean propagation values (580 km s$^{-1}$ and 370 km s$^{-1}$, respectively).

Another characteristic ICME speed of interest is the radial "centre of mass" speed or cruise speed, $v_C$, which may be defined as the speed of an ICME midway between its leading and trailing edge (Owens *et al.*, 2005). ICME 1 had an *in situ* $v_C$ value of ~394 km s$^{-1}$ at STB, very similar to the mean value during propagation, $\langle v_C \rangle = (\langle v_L \rangle + \langle v_T \rangle)/2$, of 409 km s$^{-1}$; for ICME 2, $v_C \approx 473$ km s$^{-1}$ and $\langle v_C \rangle = 475$ km s$^{-1}$. The finding that $v_C \approx \langle v_C \rangle$ indicates that the cruise speed of both ICMEs was approximately constant during propagation, in agreement with the commonly held assumption of $v_C$ constancy in ICMEs.

The *B* ratio profiles in Figures 2 and 3 may be explained in terms of the ICMEs' expansion and spacecraft crossing times. ICME 1 was likely to be expanding at the location of MES, given the shape of the field magnitude profile observed by the spacecraft (see Figure 1). When MES observed the front half of the rope, the ICME was relatively less expanded, and field magnitudes were consequently relatively high; by the time that the rear half was observed, the ICME had expanded more, and the observed magnitudes were lower. This produces the "ski ramp" profile that is characteristic of expansion (Farrugia *et al.*, 1993; Osherovich *et al.*, 1993) and which was seen in ICME 1 at MES. By the time the ICME reached STB, radial expansion across the rope had ceased, the speed profile within the rope was flat, and the field magnitude profile was symmetric (and flat). Thus, the ratio of the observed magnitudes at STB to MES is lower in the front half of the rope than in the rear half. This contrasts with ICME 2, where the ICME was expanding at both MES and STB: magnitudes in the front half were higher than in the rear half by a similar proportion at both spacecraft, hence the ratio of the two profiles is flat. The flat ratio suggests that the expansion rates of ICME 2 were similar to each other at the observing spacecraft. The greater small-scale variability in the ratio for ICME 2 than for ICME 1 is notable. The flatness in the magnitude profile for ICME 1 at STB may indicate that the flux rope's axial field component (dominant near the midpoint of the profile) dropped faster with propagation distance than the azimuthal component (dominant at the edges of the profile), an effect predicted by analytical models of flux rope evolution (*e.g.* Osherovich *et al.*, 1993; Démoulin and Dasso, 2009).

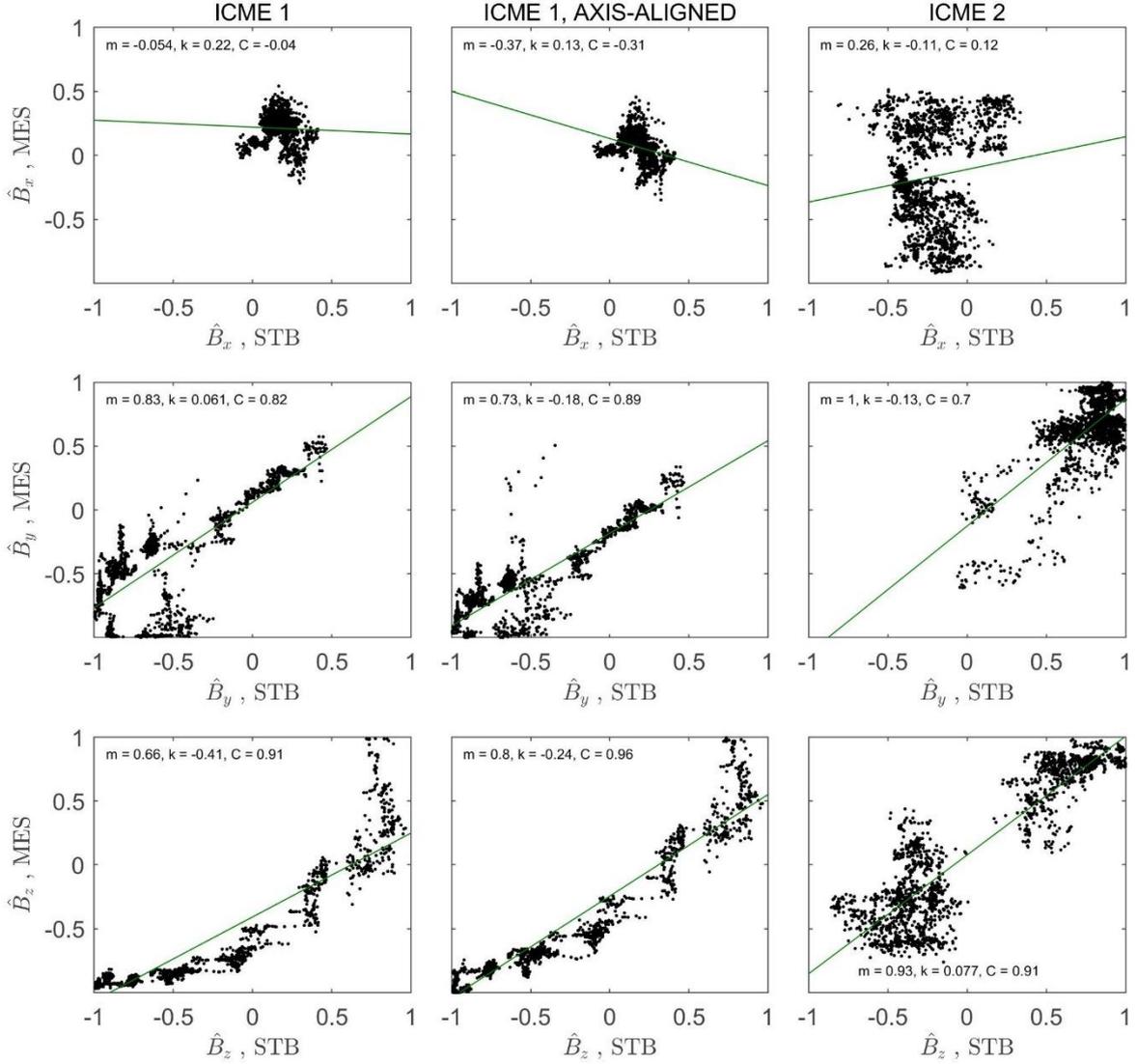

**Figure 5.** Plots of the normalised magnitudes of each field component at STB versus the normalised, interpolated magnitudes at MES, for both the ICME 1 and ICME 2 mappings. The green lines show linear least-squares fits to the data. The values of the gradient, $m$, intercept, $k$, and correlation coefficient, $C$, associated with each fit are indicated. The ICME 1 plots include 1779 vectors and the ICME 2 plots 1834 vectors.

## 2.3. Correlation of the Magnetic Field Components

### 2.3.1. Overall Correlation

Correlation coefficients have been used to determine the linearity between the mapped MES and STB profiles. Here we calculate Pearson's correlation coefficient, $C_i$, a standard measure of linear correlation that may be defined in terms of the covariance matrix,

$$C_i(\hat{B}_i, \hat{B}_{im}) = \frac{\text{cov}(\hat{B}_i, \hat{B}_{im})}{\sigma_{\hat{B}_i} \sigma_{\hat{B}_{im}}} \qquad (5)$$

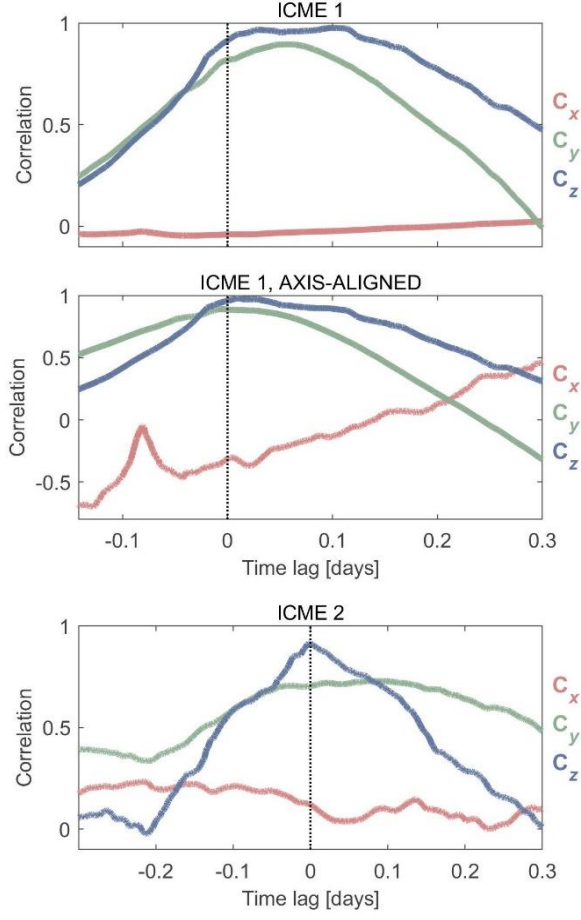

**Figure 6.** Cross-correlation for ICME 1 are shown in the top and middle panels. The MES rope data have been lagged against STB data from ~0.14 days before to 0.3 days after the rope interval at STB. The values at zero lag correspond to the correlation coefficients where the rope boundaries in the MES and STB data overlap. The cross-correlation for the single ICME 2 mapping is shown in the bottom panel.

where $\hat{B}_i$ and $\hat{B}_{im}$ denote the outer (STB) and mapped inner spacecraft (MES) magnetic field data, respectively, $\sigma_{\hat{B}_i}$ and $\sigma_{\hat{B}_{im}}$ denote their respective standard deviations, and $i = \{x, y, z\}$ denotes the field component. A correlation value of 1 indicates perfect linear correlation, a value of -1 perfect linear anti-correlation, and a value of 0 no linear correlation. Note that Pearson's correlation does not measure how much two datasets overlap, since it is invariant to origin shifts and changes in scale, *i.e.* if $B_i$ were transformed to $a + b\hat{B}_i$ and $\hat{B}_{im}$ to $c + d\hat{B}_{im}$ where $a$, $b$, $c$ and $d$ are constants, then $C_i$ would be unchanged. The mapped data have been linearly interpolated to the same resolution as the outer spacecraft data in order to determine the coefficients.

The correlation $C = [C_x, C_y, C_z]$ of the three field components for the ICME 1 mapping shown on the left-hand side of Figure 3 has values of [-0.04, 0.82, 0.91], indicating a relatively high correlation in $\hat{B}_y$ and $\hat{B}_z$, and no significant correlation in $\hat{B}_x$. The axis-aligned mapping on the right-hand side of Figure 3 has $C$ values of [-0.31, 0.89, 0.96], indicating an increased correlation in $\hat{B}_y$ and $\hat{B}_z$, and no significant change in the $\hat{B}_x$ correlation. The mapping for ICME 2 displayed in Figure 4 has a correlation of [0.12, 0.70, 0.91]: as in the ICME 1 mappings, there is a high correlation in $\hat{B}_y$ and $\hat{B}_z$, and no correlation in $\hat{B}_x$.

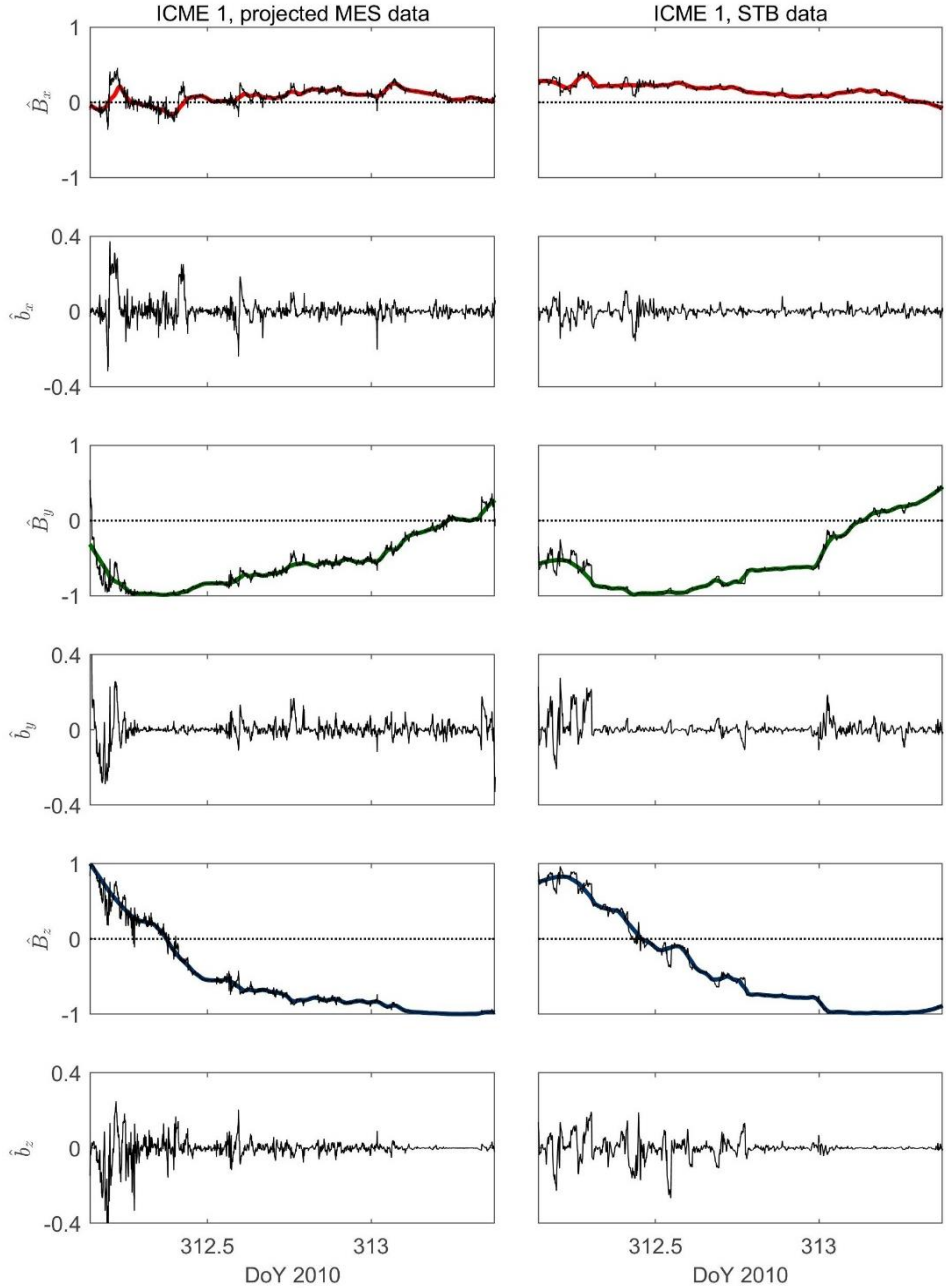

**Figure 7.** Macroscopic and microscopic structure observed in ICME 1's flux rope at MES (left panels) and STB (right panels). The smooth coloured lines show the macroscopic profiles estimated from LOWESS fitting for each component; these lines overlay the original $\hat{B}_i$ data. The microscopic profiles, $\hat{b}_i$, were obtained by subtracting the LOWESS fits from $\hat{B}_i$.

The origins of the low $C_x$ values can be seen in Figure 5. The figure shows the $\hat{B}_i$ values at STB versus the mapped, interpolated $\hat{B}_{im}$ MES values, for both the ICME 1 mappings (first and second column panels) and the ICME 2 mapping (third column). The green lines in the figure are least-squares linear fits to the data; the correlation coefficients calculated using Equation 5 reflect how well these linear fits represent the spread of data in the scatter plots. It can be seen that $\hat{B}_y$ and $\hat{B}_z$ are approximately linear for ICME 1, and somewhat less so in the ICME 2 mapping. The tight clustering of points in $\hat{B}_x$ for ICME 1 produces the low $C_x$ values previously quoted. As can be seen in Figure 3, $\hat{B}_x$ was flat and close to zero at both spacecraft, with the profiles largely in agreement with each other. The correlation values – despite the significant agreement in the time series profiles – are low because of the particularly low variance in this component. The correlation coefficient is therefore not a suitable measure of similarity for $\hat{B}_x$ in ICME 1. For ICME 2, there is a much greater spread in the $\hat{B}_x$ scatter plot in Figure 5, and the low correlation coefficient in this case does reflect the poor agreement of $\hat{B}_x$ displayed in Figure 4.

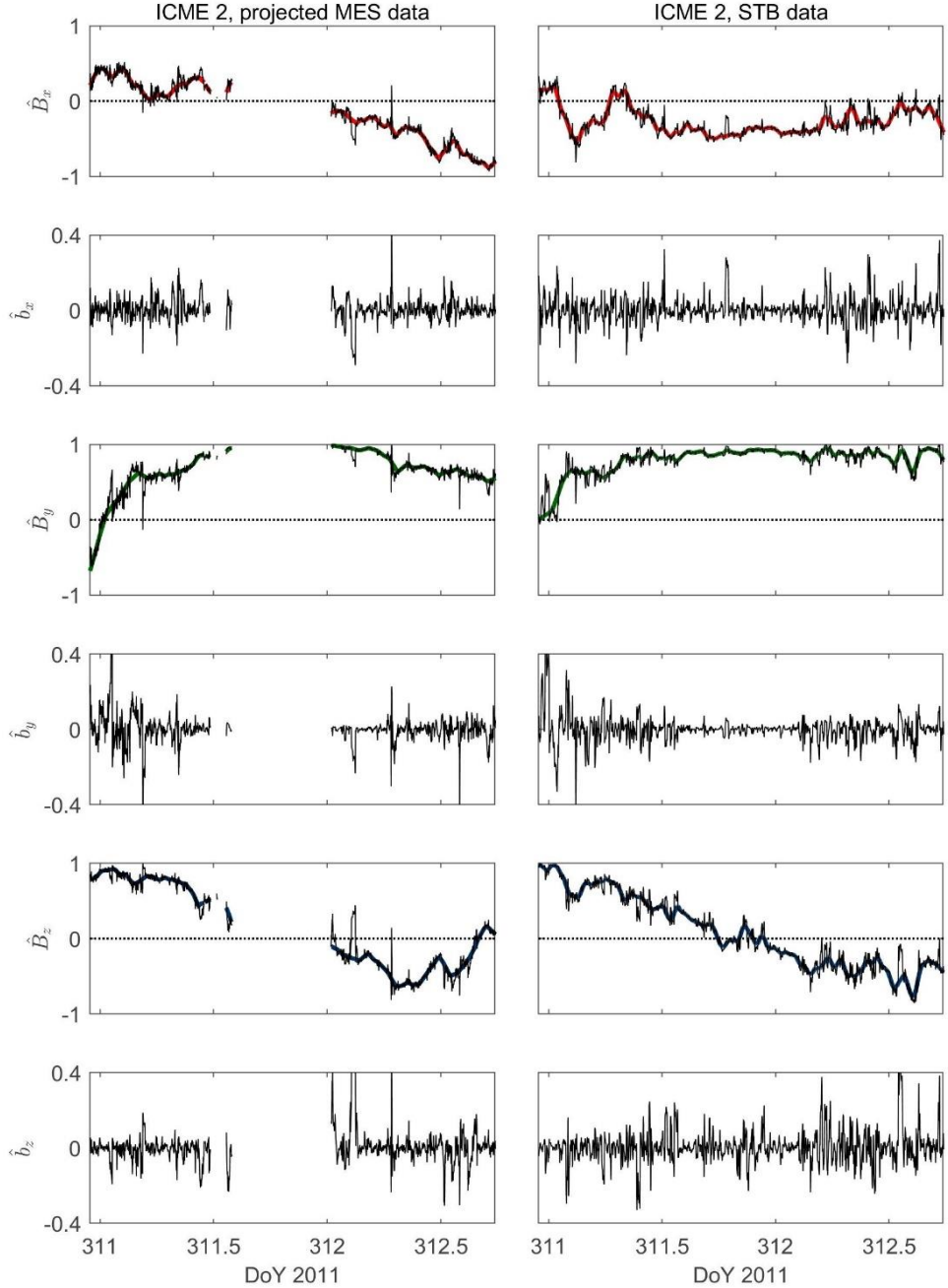

**Figure 8.** Macroscopic and microscopic structure observed in ICME 2's flux rope at MES and STB. The figure is presented in the same way as Figure 7.

The top two panels in Figure 6 show cross-correlations of the mapped MES flux rope data with STB data for ICME 1. The leading edge of the flux rope in the MES data has been lagged from

~0.14 days before to 0.3 days after the flux rope leading edge at STB. The lag is asymmetric due to a magnetic field data gap that preceded the flux rope interval at STB. The raw cross-correlation values at zero time lag have been normalised to equal the correlation coefficient values, $C$, and the other cross-correlation values have been scaled accordingly. If the expansion scaling in the mapped data is accurate and the flux rope boundaries are correctly aligned, the cross-correlation would be expected to peak at zero lag. This is not the case for the initial mapping of ICME 1, which shows a peak in $\hat{B}_y$ and $\hat{B}_z$ correlation at a lag of ~0.06 days. The axis-aligned mapping for ICME 1, in contrast, does peak at zero lag. The ICME 2 mapping also peaks in correlation at zero lag for $\hat{B}_y$ and $\hat{B}_z$, as shown in the bottom panel of Figure 6. The leading edge of the MES flux rope data was lagged between 0.3 days before and after the leading edge in the STB data to produce this figure. $\hat{B}_x$ shows no significant correlation at any lag for either event.

### 2.3.2. Correlation at Different Temporal Scales

The $C$ coefficients above give the overall correlation across all temporal scales. Features within the same component that vary with different characteristic timescales may not show the same degree of correlation. Here we define macroscopic features in the profiles to be those that vary with a timescale of 1 hour, and microscopic features to be those that vary with a timescale of 1 minute. This choice of timescales is somewhat arbitrary, but does allow the relationship between correlation and timescale to be illustrated effectively. Shocks, tangential discontinuities and reconnection exhausts are typically observed at timescales of ~1 minute in the solar wind at 1 AU, while large-scale heliospheric structures such as ICMEs vary at timescales of ~1 hour.

To obtain profiles of macroscopic features, robust LOWESS smoothing (Cleveland, 1979) has been applied to the magnetic field data. This smoothing is similar in effect to a low-pass filter. In outline, the LOWESS technique involves selecting a span of data centred on the point to be smoothed, determining weights for each point within the span, and performing a weighted linear least-squares regression with a first-order polynomial across the span. The robust version of the technique applies an additional weighting, where points with high residual values relative to the initial regression are reduced in weight. The regression is then recalculated with the additional weighting. The additional weighting and regression are performed iteratively five times until a final regression is obtained; the value of the final regression at the point of interest gives the smoothed value. These steps are repeated for all points in the data series. A span width of 1 hour was used. Further details on the LOWESS technique used may be found at `mathworks.com/help/curvefit/smoothing-data.html`. Profiles of the microscale features were obtained by subtracting the smoothed values from the original datasets.

The macroscale correlation, $C_L$, in the initial and axis-aligned mappings of ICME 1 were found to be [-0.02, 0.85, 0.92] and [-0.41, 0.94, 0.96], respectively, indicating strong $\hat{B}_y$ and $\hat{B}_z$ macroscale correlations, and increased correlations relative to the overall values. The $\hat{B}_x$ correlations are again low because of the low variance found in this component. The corresponding microscale correlation, $C_S$, for the two mappings were [-0.06, 0.05, 0.13] and [-0.09, 0.06, 0.12], indicating that there was no correlation of microscale features in any component for either mapping. The ICME 2 correlations, $C_L = $ [0.14, 0.84, 0.94] and $C_S = $ [0.03, 0.02, -0.07], show a similar trend to that found in the ICME 1 correlations.

Figure 7 and 8 show the macroscale and microscale profiles for each component in ICME 1 and ICME 2, respectively, at both MES and STB. The smooth coloured lines overlaying the $\hat{B}_i$ profiles represent the macroscale structure estimated from the LOWESS fitting, and the $\hat{b}_i$ panels show the microscale structure. It can be seen that more microscale structure was observed in ICME 2 than in ICME 1, and that, despite the low $C_S$ values, some microscale features appear to have been retained during propagation (*e.g.* a microscale feature near the leading edge of ICME 2 in $\hat{b}_y$, at around DoY 311.1).

## 3. Discussion

The mappings displayed in Figures 3 and 4 indicate that there was a high degree of similarity between the ICME magnetic field profiles observed at MES and STB. The correlation analysis derived from these mappings gives a good quantified measure of the similarity in the solar east-west and north-south directions: features at the macroscopic scale in the two events studied were well correlated, while features at the microscopic scale were uncorrelated. Macroscale features broadly correspond to the underlying flux rope structure of the ICMEs. The degree of similarity observed is remarkable given the ~2-day propagation time and ~0.6 AU propagation distance for both ICMEs.

Others who have studied ICMEs observed at spacecraft with very small longitudinal separations (*e.g.* Mulligan *et al.*, 1999; Nackwacki *et al.*, 2011) have reported comparable levels of similarity in macroscopic field structure with propagation distance for some ICMEs. The approximate self-similarity in the field profiles is also in agreement with the analytical flux rope modelling of Démoulin and Dasso (2009). Their work indicates that a range of flux rope configurations would expand in the radial direction almost self-similarly with heliocentric distance, $r_H$, if the ropes are embedded in solar wind plasma with pressure that falls according to an empirically derived $r_H^{-2.8}$ power law (*e.g.* Gazis *et al.*, 2006).

The field profile similarity may be interpreted in two ways. If the same region of each ICME was sampled by the observing spacecraft pair – *i.e.* if the spacecraft angular separations can be neglected – then the similarity suggests there was little evolution in the observed region during propagation, and that the field structure of the observed region was robust. Alternatively, if adjacent regions were observed by the two spacecraft – *i.e.* if the spacecraft angular separations cannot be neglected – then the similarity indicates that the adjacent regions were similar, and that there was spatial homogeneity across the angular extent sampled. Different regions may also have been sampled if there had been significant non-radial components to the ICMEs' propagation velocities.

Determining which of these interpretations is correct is difficult given the observations available. It may be possible to neglect the spacecraft angular separations if they were small relative to the overall latitudinal and longitudinal extents of the ICMEs; however, these global extents cannot be determined from the *in situ* measurements analysed. Compared to the average CME latitudinal span of 50° to 60° seen in coronagraph images (*e.g.* Yashiro *et al.*, 2004), the spacecraft longitudinal separations (1.0° for ICME 1 and 3.5° for ICME 2) were indeed relatively small, whereas the latitudinal separations (7.0° for ICME 1 and 6.8° for ICME 2) were somewhat more significant.

In ICME 1, the overall increase in profile similarity and correlation of the axis-aligned fields relative to the initial, unaligned mapping is notable. If the two spacecraft observed the same region of the ICME, the increased correlation indicates that the local flux rope orientation changed during propagation. The increased correlation would suggest that the MVA-determined axis directions are reasonably accurate. Given how well the axis-aligned profiles overlap all along their length, it would also suggest that this apparent rotation did not distort the underlying structure of the flux rope.

It has been assumed that reconnection did not erode the ICME flux ropes during propagation between the spacecraft by any significant amount. The strong similarities seen in the profiles, and the agreement in field direction at the leading and trailing edges, would support this assumption. Many ICME flux ropes arriving at 1 AU show signs of erosion (Ruffenach *et al.*, 2015), but much of this erosion is thought to occur within the orbital radius of Mercury (Lavraud *et al.*, 2014) where the Alfvén speed, and hence the reconnection rate, is considerably higher. If a significant amount of erosion had occurred at the rope edges during propagation, points within the rope interval (*i.e.* not the boundaries) at the inner spacecraft would need to be mapped to the boundaries observed at the outer spacecraft. The upcoming *Solar Orbiter* (Müller *et al.*, 2013) and *Parker Solar Probe* (Fox *et al.*, 2016) missions will travel to within the orbit of Mercury, where ICMEs may appear less eroded than at 1 AU.

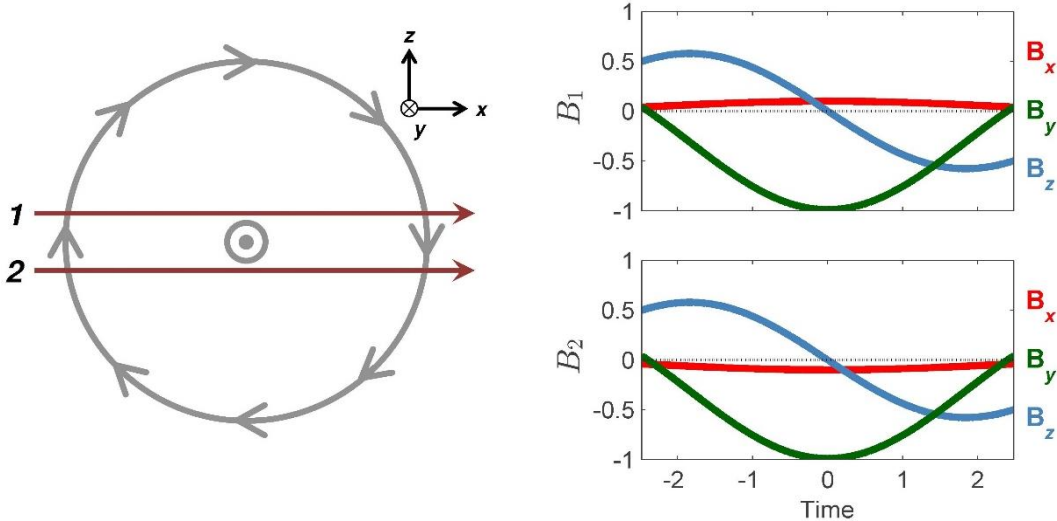

**Figure 9.** An idealised flux rope with a circular cross section is displayed in the left-hand panel. Two different trajectories by spacecraft through the flux rope, labelled 1 and 2, are shown. The axis of the rope is aligned with the $-y$ direction. Lundquist profiles that would be observed by spacecraft following trajectories 1 and 2 are displayed on the right-hand side. $B_y$ and $B_z$ are perfectly correlated, $B_x$ perfectly anti-correlated. Zero time in the right-hand panels is the time of closest approach to the rope axis by the spacecraft; units are arbitrary.

The nature of the microscale features displayed in Figure 7 and 8 is not considered in this work. However, we speculate that these features may be substructures that are smaller in angular extent than the angular spacecraft separation (and hence not encountered by both spacecraft), or they may be structures that evolve significantly during propagation. They may also be temporal, transient features such as waves, the observation of which is dependent on the local wave speed. Features of these kinds would not be correlated at the different spacecraft.

The correlation is of course sensitive to how the mapping is performed, and it would be worthwhile to consider whether other mappings would produce an increase in the correlation at microscopic scales. Other mappings could involve removing the assumption of linearity in the mean rope velocity profile, for example, or relaxing the condition that the predetermined rope edges at each spacecraft must line up with each other. One could attempt a least-squares mapping that minimises the residuals between the datasets, where axis directions, rope boundaries and the velocity profile are free parameters.

There are limitations to the correlation analysis technique presented in Section 3.2. It does not measure overall similarity of the vectors, and is only suited to a component-by-component analysis. In the case of ICME 1, the similarity seen in the $\hat{B}_x$ profiles is not reflected in the correlation coefficient, due to the low variance of this component. However, for both ICMEs, the high correlation values for $\hat{B}_y$ and $\hat{B}_z$ do reflect, and give a quantified measure, of the high degree of similarity seen in these components. We intend in future studies to develop the correlation analysis technique introduced here.

There are some noticeable anti-correlated features in the $\hat{B}_x$ profiles for ICME 2, possibly a result of different flux rope axis directions in the two datasets. However, these features could be due to the spacecraft taking slightly different trajectories through the ICME. Figure 9 shows an idealised flux rope with a circular cross-section on the left-hand side, with trajectories by two spacecraft through the rope. The flux rope axis points in the $-y$ direction, and the trajectories are in the $x$ direction. Trajectory 1 crosses the rope just above the axis, and Trajectory 2 just below; the impact parameters of the trajectories have equal magnitudes. The panels on the right-hand side show the force-free Lundquist magnetic field profiles (*e.g.* Burlaga, 1988) that would be observed by the two spacecraft during the passage of the idealised flux rope; the $\hat{B}_y$ and $\hat{B}_z$ profiles are perfectly correlated, and the $\hat{B}_x$ profiles perfectly anti-correlated. Thus, a small deviation from radial alignment that results in the spacecraft traversing the rope either side of the axis could produce the correlations observed in ICME 2. However, it cannot be determined whether the scenario described above arose for this ICME without further analysis of its global structure and the spacecraft impact parameters.

From a space weather perspective, the macroscale structure of an ICME is more significant than any short-duration microscale features. ICMEs are a major source of sustained periods of southward-directed magnetic field at 1 AU. There is now much focus in the scientific community on attempting to forecast $B_z$ in ICMEs incident at the Earth with forecast lead times greater than the ~45 minutes provided by ACE, *Wind* and DSCOVR at L1. Many current efforts concentrate on using remote observations to make predictions (*e.g.* Savani *et al.*, 2015; Möstl *et al.*, 2017), although some have proposed *in situ* space weather monitors that could be placed well upstream of the Earth (*e.g.* the *Sunjammer* mission concept; Eastwood *et al.*, 2015); Kubicka *et al.* (2016) have recently proposed a method to forecast $B_z$ that combines remote and *in situ* observations. If ICME magnetic field structure in the inner heliosphere is generally well correlated along radial lines from the Sun, as in the case of the two ICMEs studied in this work, then the task of predicting ICME magnetic field properties at 1 AU from sub-1 AU *in situ* observations would be made less difficult: a model that accurately predicts arrival times, radial expansion and magnitude from sub-1 AU observations would suffice. However, if the underlying magnetic topologies of ICMEs are significantly altered during propagation through strong interactions with solar wind structures, as in a case reported by Winslow *et al.* (2016), or through interactions with other ICMEs (Lugaz *et al.*, 2016, and references therein) then more complex modelling would be required. How commonly such changes in ICME field structure occur remains an open question. Statistical analyses of more ICMEs observed by radially aligned spacecraft pairs are needed to shed light on this matter.

## 4. Conclusion

The magnetic flux rope profiles of two ICMEs each observed by a pair of spacecraft near radial alignment have been analysed. The spacecraft were separated by more than 0.6 AU in heliocentric distance, less than 4° in heliographic longitude, and 7° in heliographic latitude during the observation period. Rope data from the inner spacecraft have been mapped to the outer spacecraft by using mean, linearly declining speed profiles estimated from arrival times. Figures that result from this mapping allow for easy qualitative assessment of the similarity in field structure, and reveal similarities in the datasets not readily apparent in the direct observations (Figure 1). Correlation coefficients for each field component have also been determined, at both micro- and macroscopic temporal scales. In the ICMEs analysed, it has been found that:

i) Both ICMEs expanded in the radial direction during propagation, and both propagated at approximately constant centre-of-mass cruise speeds;

ii) There was a high degree of qualitative similarity in the flux rope profiles;

iii) Macroscale features in the profiles, which correspond approximately to the underlying flux rope structure, were well correlated in the $y$ and $z$ directions, with Pearson's correlation coefficients of ~0.85 and ~0.95 respectively;

iv) Microscale features that varied on timescales of ~1 minute were uncorrelated;

v) Macroscale correlation in one of the ICMEs analysed was increased when an apparent rotation of 19° by the flux rope axis was considered;

vi) The similarity in the field profiles may be interpreted in two ways. If the same region of each ICME was intersected by the observing spacecraft, it indicates that the underlying, large-scale *B*-field structure of the observed regions remained intact during propagation. If the same region was not observed in each case, it indicates homogeneity in field structure across the angular extent of the ICME spanned by the spacecraft.

If the similarity in magnetic field structure at different heliocentric distances seen in these ICMEs is common, then the task of predicting $B_z$ in ICMEs arriving at the Earth using an upstream, *in situ* space weather monitor would be much simplified.

**Acknowledgements**

We wish to thank the MESSENGER and STEREO instrument teams for providing the data used in this work, and the PDS:PPI and SPDF CDAWeb data archives for their distribution of data. We also wish to thank the referees for their comments and suggestions for improvements to the manuscript. This work has been supported with funding provided by the European Union's Seventh Framework Programme for research, technological development and demonstration under grant agreement No. 606692 [HELCATS].

**Disclosure of Potential Conflicts of Interest**

The authors declare that they have no conflicts of interest.